\documentstyle[12pt]{article}

\begin{document}

\begin{center}
{\Large{\bf On the structure of exact effective action for N=1
supersymmetric theories}}

\vspace{0.5cm}

{\large K.Stepanyantz}%
\footnote{E-mail:$stepan@theor.phys.msu.su$}

\end{center}

\begin{center}
{\em Moscow State University, physical faculty,\\
department of theoretical physics.\\
$117234$, Moscow, Russia}
\end{center}

\begin{abstract}
We discuss the ways of constructing the exact superpotential for N=1
supersymmetric theories and propose a new approach. As a consequence,
a new structure of the superpotential is found.
\end{abstract}


\section{Introduction}
\hspace{\parindent}

The existence of N=1 supersymmetry in the standard model was confirmed
by the indirect experimental data \cite{exper,dienes}. That is why the
dynamics of supersymmetric theories should be thoroughly investigated,
especially beyond the frames of the perturbation theory. This case, being
the most complicated, is of the most interest, because nonperturbative
effects seem to produce the quark confinement.

Recently there is a considerable progress in understanding of the
nonperturbative dynamics, caused by \cite{seiberg}. In this paper the sum
of instanton corrections was found explicitly for the simplest N=2
supersymmetric Yang-Mills theory with $SU(2)$ gauge group. However, this
model is very far from the realistic ones, the investigation of N=1
supersymmetric models being much more interesting.

N=1 supersymmetric theories were considered first in \cite{affleck}. In
this paper the effective potential was found from the exact conservation
of R-symmetry current beyond the perturbation theory. This approach was
developed in \cite{seiberg2}, where it was proposed to use composite
fields for the description of the theory for $N_f\ge N_c$.

However, the corresponding superpotential does not agree with instanton
calculations and does not reproduce quantum anomalies. These problems can
be solved by introducing a new composite field -- gluino condensate. At
the perturbative level anomalies are reproduced by the
Veneziano-Yankielowitch effective Lagrangian \cite{veneziano}. However, it
is not applicable beyond the frames of the perturbative theory. At the
nonperturbative level it is possible to use the relation between
perturbative and exact anomalies, which allows to construct a
superpotential, that is in agreement with the transformation law of the
collective coordinate measure and reproduces quantum anomalies \cite{hep}.
For $N_c \le N_f$ it is possible to integrate the gluino condensate out of
it and obtain Seiberg's exact results, but for $N_c > N_f$ it is not so
\cite{1}.

However, in the present paper we argue, that the method, based on using
of composite fields, suffers from some problems and propose a new approach.
As a consequence we find a new form of the effective superpotential.

The paper is organized as follows: In Section \ref{rel} we briefly
remind some results, obtained in the frames of the approach, proposed by
Seiberg -- Affleck-Dine-Seberg superpotential (section \ref{aff}) and
nonperturbative generalization of Veneziano-Yankielowitch effective
Lagrangian (section \ref{ven}). Here we discuss main shortcomings of these
expressions and make a conclusion, that a new approach is needed. This new
approach is formulated in Section \ref{approach}. In this section we
also construct our main result -- exact effective action for the massive
N=1 supersymmetric Yang-Mills theory with $SU(N_c)$ gauge group and $N_f$
matter supermultiplets. The results are briefly discussed in the Conclusion.
In the Appendix we derive the expression for the superpotential and prove
its uniqueness.


\section{Exact superpotential -- usual approach}
\label{rel}

\subsection{Affleck-Dine-Seiberg superpotential}
\hspace{\parindent}
\label{aff}

The first attempt to construct the exact superpotential for $N=1$
supersymmetric theories was made in \cite{affleck}.

It is well known, that in this case there is a special chiral symmetry
(so called R-symmetry), which is not destroyed by the perturbative
quantum corrections. Assuming, that the correspondent current is also
conserved beyond the frames of the perturbation theory, Affleck, Dine
and Seiberg constructed the following superpotential, depending only
on scalar fields and reproducing zero anomaly of R-symmetry:

\begin{equation}\label{ads}
L_a = \int d^2\theta \Lambda^{\frac{3 N_c-N_f}{N_c-N_f}}
\phi^{- \frac{2 N_f}{N_c-N_f}}
\end{equation}

\noindent
(Here $L_a$ denotes the holomorphic part of the effective Lagrangian.)

However, this superpotential can not be generated by instantons, because
for the N=1 supersymmetric Yang-Mills theory with $SU(N_c)$ gauge group
and $N_f$ matter supermultiplets the instanton corrections should be
proportional to

\begin{equation}
L_a \sim \Lambda^{(3 N_c - N_f) n}
\end{equation}

\noindent
where $n$ is the module of a topological number.

Moreover, (\ref{ads}) does not reproduce anomalies (except for R-symmetry)
according to the equation

\begin{equation}
\langle \partial_\mu J^\mu_{(\alpha)}\rangle
= - \frac{\partial\Gamma}{\partial\alpha}
\end{equation}

\noindent
because it does not contain gauge degrees of freedom.

The development of the ADS-approach was given in \cite{seiberg2}. The
dynamics was argued to depend crucially on the numbers of colors and
flavors. For example, for $N_f>N_c$ it was proposed to use gauge invariant
variables, parametrizing the moduli space, as new quantum fields, while
for $N_f<N_c$ the effective action depends on the original fields. For
$N_f=N_c$ instanton corrections to a special classical constrain lead to
the necessity of introducing Lagrange multiplier to the effective action.

However, the problems mentioned above were not solved. Moreover, these
results do not explain the phenomenon of confinement.


\subsection{Nonperturbative generalization of Veneziano-Yankielowitch
effective Lagrangian}
\hspace{\parindent}
\label{ven}

The development of this approach in \cite{hep} allowed to construct a
superpotential, agreeing with the transformation law of the collective
coordinate measure and reproducing anomalies of chiral symmetries. The
matter is that constructing exact superpotential it is impossible to omit
gauge degrees of freedom. In the frames of the above approach it is a
gluino condensate $S=W_a^2$.

Taking this dependence into account it is possible to find the following
expression for the superpotential \cite{hep}

\begin{equation}\label{result}
L_a = \frac{1}{16\pi} \mbox{Im} \int d^2\theta\ S \tau(z^{-1/4})
\end{equation}

\noindent
Here $\tau(a)$ coincides with the correspondent function in the
Seiberg-Witten solution, defined by

\begin{equation}\label{tau}
\tau(a) = \left. \frac{da_D(u)}{da}\right|_{\displaystyle u=u(a)}
\end{equation}

\noindent
where

\begin{equation}\label{sw}
a(u)=\frac{\sqrt{2}}{\pi}\int\limits_{-1}^{1} dx
\frac{\sqrt{x-u}}{\sqrt{x^2-1}};
\qquad
a_D(u)=\frac{\sqrt{2}}{\pi}\int\limits_{1}^{u} dx
\frac{\sqrt{x-u}}{\sqrt{x^2-1}}.
\end{equation}

In the frames of the approach, proposed by Seiberg, the parameter $z$ is
a function of "mesons", "barions" and gluino condensate $S$. The exact
expression, found in \cite{hep}, has the following form:

\begin{eqnarray}\label{z1}
&&z=\frac{\Lambda^{3N_c-N_f}}{\mbox{det} M\ S^{N_c-N_f}},\qquad N_c > N_f
\nonumber\\
&&z=\frac{\Lambda^{3N_c-N_f}S^{N_f-N_c}}{
\mbox{det} M - (\tilde B^{A_1 A_2\ldots A_{N_f-N_c}} M_{A_1}{}^{B_1}
\ldots M_{A_{N_f-N_c}}{}^{B_{N_f-N_c}}
B_{B_1 B_2\ldots B_{N_f-N_c}})},\nonumber\\
&&N_c \le N_f
\end{eqnarray}

Taking the asymptotic of the exact solution

\begin{equation}
\tau(a) \to \frac{2i}{\pi} \ln a, \qquad a\to \infty \quad (z \to 0)
\end{equation}

\noindent
we obtain, that at the perturbative level (\ref{result}) coincides with
the Veneziano-Yankielowitch effective Lagrangian.

Therefore, (\ref{result}) can be considered as a synthesis of Seiberg's
exact results and Veneziano-Yankielowitch effective Lagrangian.

In addition we should point out, that the gluino condensate $S$ is a
natural Lagrange multiplier if $N_f=N_c$ and, moreover, (\ref{result})
allows to treat theories with N=2 and N=1 supersymmetry in a similar way.

However, (\ref{result}) (with the parameter $z$ given by (\ref{z1}))
suffers from some problems. For example, it seems, that a theory can not
be described by gauge invariant composite fields in principle. Really,
note, that in the frames of the above approach the gluino condensate
should be considered as a new scalar quantum field, because otherwise some
operations with $S$ become senseless. For example, if $S=W_a^2$ it is
impossible to take $\ln S$, because the lowest component of this
superfield contains anticommuting spinors. Moreover, all sufficiently
large powers of $S$ are simply equal to 0 due to the same reason.
Therefore, $S$ is a new scalar field which should be integrated out on
shell \cite{intriligator}. If it is possible, formally we should obtain
Seiberg's exact results. Nevertheless, it is impossible for $N_c>N_f$,
because in this case the equation

\begin{equation}
\frac{\partial w}{\partial S} = 0
\end{equation}

\noindent
has no solutions \cite{1}. For $N_c\le N_f$ there are solutions, but, as
we mentioned above, integrating the gluino condensate out of the effective
action seems to be impossible in principle. By other words, the Yang-Mills
theory should be formulated in terms of the gauge field $A_\mu^a$, but not
in terms of gauge invariant $F_{\mu\nu}^2$, and, therefore, $S$, $M_A^B$
and so on can not be considered as quantum fields.


\section{Another approach to construct exact superpotential}
\hspace{\parindent}
\label{approach}

Taking into account the arguments, given in the previous section, it is
necessary to propose another approach for constructing an exact
superpotential. An exact result should satisfy the following evident
conditions:

{\bf 1.} It should depend on the original fields of the theory.

{\bf 2.} It should agree with dynamical (perturbative and instanton)
calculations (This requirement is much more restrictive, than the
agreement with the transformation law of the collective coordinate
measure).

The purpose of the present paper is to construct an expression for the
effective superpotential, satisfying these requirements.

First, let us find its general structure. Note, that there is a relation
between perturbative and instanton contributions
\cite{thooft,shifman_beta,shifman_review}. For example, the
renorminvariance of instanton corrections allows to construct exact
$\beta$-functions of supersymmetric theories.

Let us express this relation mathematically. Chose a scale $M$ and
denote the value of the coupling constant at this scale by $e$. Then
the perturbative result is proportional to $-1/4e^2$, while the instanton
contributions are proportional to $\exp(-8\pi^2 n/e^2)$, where $n$ is the
module of a topological number. (Perturbative and (each of) instanton
contributions are renorminvariant separately.)

The holomorphic part of the {\it perturbative} effective action can be
written as

\begin{equation}
L_a = \frac{1}{16\pi}\mbox{Im}\,\mbox{tr}
\int d^2\theta\,W^2
\Bigg(\frac{4\pi i}{e^2_{eff}}+\frac{\vartheta_{eff}}{2\pi}\Bigg)
\end{equation}

\noindent
where $e_{eff}$ and $\vartheta_{eff}$ are renorminvariant functions of
fields, the effective {\it perturbative} coupling and $\vartheta$-term
respectively. Denoting

\begin{equation}
z \equiv \exp\Bigg(2\pi i
\Big(\frac{4\pi i}{e^2_{eff}}+\frac{\vartheta_{eff}}{2\pi}\Big)\Bigg)
\end{equation}

\noindent
we obtain, that the exact effective action (with instanton contributions)
can be written as

\begin{equation}\label{gs}
L_a = -\frac{1}{32\pi^2}\mbox{Re}\,\mbox{tr}\int d^2\theta\ W^2 f(z)
= -\frac{1}{32\pi^2}\mbox{Re}\,\mbox{tr}\int d^2\theta\ W^2 \Big(\ln z
+\sum\limits_{n=1}^\infty c_n z^n\Big)
\end{equation}

Note, that this expression is valid beyond the constant field approximation
(that is usually assumed in derivation of the exact results).

The function $f(z)$ can be found explicitly. The matter is that the
conditions

\begin{equation}
f(z) = \ln z + \sum\limits_{k=1}^\infty c_n z^n;\qquad
\mbox{Re}\,f(z)<0
\end{equation}

\noindent
define its form uniquely. (The latter equation is a requirement of
a positiveness of the squared effective coupling). In the Appendix
\ref{fform} we prove, that these conditions uniquely lead to

\begin{equation}
f(z) = 2\pi i\, \tau(z^{-1/4})
\end{equation}

\noindent
where $\tau(a)$ is the Seiberg-Witten solution (\ref{tau}). Therefore,
finally,

\begin{equation}\label{result2}
L_a = \frac{1}{16\pi} \mbox{Im} \int d^2\theta\, W^2 \tau(z^{-1/4})
\end{equation}

\noindent
where the parameter $z$ can be found approximately by the one-loop
calculations and exactly (up to a constant factor) by investigation of
the collective coordinate measure, similar to
\cite{shifman_beta,shifman_review}.

Let us consider, for example, N=1 supersymmetric Yang-Mills theory
with $SU(N_c)$ gauge group and $N_f$ matter supermultiplets. Because
in the one-loop approximation

\begin{equation}
\beta(e)= - \frac{e^3}{16\pi^2} (3N_c-N_f)
\end{equation}

\noindent
the parameter $z$ should be proportional to $M^{3N_c-N_f}$, where $M$ is
an UV-cutoff. This result, of course, can be obtained from the collective
coordinate measure \cite{cordes}.

The exact results, reminded in the previous section, were obtained for a
massless case. However, in realistic theories most fields are massive
(except for the gauge bosons, corresponding to an unbroken group). That is
why the massive case is the most interesting for the physical applications.
Of course, we will assume, that the supersymmetry is broken, although
will not care about the concrete mechanism. It is important only, that
there are some soft terms in the Lagrangian.

The purpose is to construct the effective action in the low-energy limit,
below the thresholds for all massive particles. The contributions of
massive particles into the running coupling constant are fixed at the
masses and, therefore, their contributions to the parameter $z$ will be
proportional to $M/m$ in a power, defined by the corresponding coefficient
of the $\beta$-function. It is much more difficult to investigate the
contributions of massless gauge fields. Of course, in this case the
coupling constant is not fixed at a definite value and we need to perform
a detailed analysis of the IR behavior of the theory. Note, that we are
interested not in the renormgroup functions but in the effective action,
that can be calculated, for example, in the constant field limit.
Therefore, the contribution of the massless gauge field to the parameter
$z$ will be a function of these fields.

Because the parameter $z$ is a scalar, we need to find a scalar superfield
containing $F_{\mu\nu}$ and not having anticommuting fields in the lowest
component (otherwise all sufficiently large powers of $z$ will be equal to
0 or infinity). The only such superfield is

\begin{equation}
B = -\frac{1}{8} \bar D (1-\gamma_5) D (W_a^{*})^2
= (D^a)^2 -\frac{1}{2}(F_{\mu\nu}^a)^2
-\frac{i}{2} F_{\mu\nu}^a \tilde F_{\mu\nu}^a + O(\theta)
\end{equation}

\noindent
where the index $a$ runs over the generators of a gauge group. (At the
perturbative level similar expression was proposed in \cite{shifman_d}).

Therefore, taking into account dimensional arguments, we obtain, that
the contribution of massless gauge fields to the parameter $z$ is
proportional to $M/B^{1/4}$ in a power, defined by the corresponding
coefficient of $\beta$-function.

The $\beta$-function of the considered model can be written as

\begin{equation}
\beta(e) = -\frac{e^3}{16\pi^2} (c_{gauge}+c_\lambda +c_q+c_{sq})
\end{equation}

\noindent
where

\begin{equation}
c_{gauge}=\frac{11}{3} N_c;\qquad c_\lambda = -\frac{2}{3}N_c;\qquad
c_q = -\frac{2}{3}N_f;\qquad c_{sq}= -\frac{1}{3} N_f
\end{equation}

\noindent
are contributions of gauge fields (with ghosts), their spinor superpartners,
quarks and squarks respectively. So, according to the above arguments, we
obtain, that

\begin{equation}
z = e^{-8\pi^2/e^2} M^{3N_c-N_f}
\Bigg(\frac{m_\lambda^{2/3}}{B^{11/12}}\Bigg)^{N_c}
\Big(m_q^{2/3} m_{sq}^{1/3}\Big)^{N_f}
\end{equation}

\noindent
where $m_\lambda$ is a gluino mass, $(m_q)_i^j$ is a quark mass matrix
and $(m_{sq})_i^j$ is a squark mass matrix.

However, this expression was found in the frames of the one-loop
approximation. To take into account multiloop effects we should note
\cite{shifman_beta,shifman_review}, that the collective coordinate
measure \cite{cordes} contains a factor

\begin{equation}
\Bigg(\frac{8\pi^2}{e^2}\Bigg)^{N_c}
\end{equation}

\noindent
Of course, it will be also present in $z$. So, finally, the parameter $z$
(up to a constant factor $C$) is written as

\begin{equation}\label{correct_z}
z = C \Bigg(\frac{8\pi^2}{e^2}\Bigg)^{N_c}
e^{-8\pi^2/e^2} M^{3N_c-N_f}
\Bigg(\frac{m_\lambda^{2/3}}{B^{11/12}}\Bigg)^{N_c}
\Big(\mbox{det}(m_q)_i^j\Big)^{2/3}
\Big(\mbox{det}(m_{sq})_i^j\Big)^{1/3}
\end{equation}

\noindent
where we take into account $SU(N_f)$ symmetry under the global rotations
in the flavor space.


\section{Conclusion}
\hspace{\parindent}

The main result of the present paper is the effective superpotential
(\ref{result2}) with the parameter $z$, given by (\ref{correct_z}). This
expression is in agreement with the perturbative calculations and possibly
do not contradict to instanton calculations (although the explicit check,
similar to \cite{yung} seems to be very interesting). Moreover, it is easy
to see, that this result is in agreement with the exact $\beta$-function,
calculated in \cite{shifman_beta}.

The structure (\ref{result2}) is different from the exact results, found
in \cite{hep}, and, especially, in \cite{affleck} or \cite{seiberg2}. The
results depend on the original (instead of composite) fields. The presence
of mass is also very important. In the massless limit the obtained results
become ill defined. It is very difficult to say, if it is necessary to
consider massless theories. At least, these theories have complicated
problems in the IR-region.

In the present paper we did not discuss the physical consequences of the
results. It will be done separately in other papers.


\vspace{1cm}

\noindent
{\Large\bf Acknowledgments}

\vspace{0.5cm}

The author is very grateful to professor P.I.Pronin for valuable
discussions and especially likes to thank V.V.Asadov for the financial
support.


\vspace{1cm}

\appendix

\noindent
{\Large {\bf Appendix}}


\section{Explicit expression for $f(z)$ and its uniqueness}
\hspace{\parindent}
\label{fform}

Let us prove, that the conditions

\begin{equation}\label{stru}
f(z) = \ln z + \sum\limits_{k=1}^\infty c_n z^n;\qquad
\mbox{Re}\,f(z)<0
\end{equation}

\noindent
uniquely define the function $f(z)$ and find it explicitly. For this
purpose we introduce a new variable $a\equiv z^{-1/4}$ (this choice of
the power will be explained below) and write formally

\begin{equation}\label{afa}
a \frac{df}{da} = - 4\frac{d^2 u}{da^2}
\end{equation}

\noindent
where $u(a)$ is an undefined function. Defining $F(a)$ as

\begin{equation}\label{F_def}
2\pi i \frac{d^2 F}{da^2} \equiv f,
\end{equation}

\noindent
we can rewrite (\ref{afa}) in the form, similar to the exact anomaly in
the Seiberg-Witten model \cite{matone,howe}

\begin{equation}\label{anom}
F + F_D = - \frac{2i}{\pi} u
\end{equation}

\noindent
where

\begin{equation}
F_D = F - a a_D; \qquad a_D = \frac{dF}{da}
\end{equation}

\noindent
Differentiating (\ref{anom}) with respect to $u$ twice, we obtain, that

\begin{equation}\label{wr}
a_D \frac{d^2a}{du^2} - a \frac{d^2a_D}{du^2} = 0
\end{equation}

Therefore, the functions $a(u)$ and $a_D(u)$ can be identified with two
linear independent solutions of the equation

\begin{equation}\label{pf}
\Big(\frac{d^2}{du^2} + L(u)\Big)\left(
\begin{array}{cc}
a\\a_D
\end{array}
\right)
\end{equation}

\noindent
where $L(u)$ is an (so far) undefined function. Substituting the asymptotic
$f_{pert}(z) = \ln z$ into (\ref{afa}), we obtain, that $u_{pert}=a^2/2$.
Therefore, in the frames of perturbation theory

\begin{equation}
L(u)_{pert} = \frac{1}{4u^2}
\end{equation}

Using (\ref{F_def}) we find, that

\begin{equation}
(a_D)_{pert} = \frac{2i}{\pi} \sqrt{2u} \Big(\ln \sqrt{2u} - 1\Big)
\end{equation}

\noindent
is a second linear independent solution of this equation. Hence, the
monodromy at the infinity is

\begin{equation}
M_\infty =
\left(
\begin{array}{rr}
-1 & 0\\
2 & -1
\end{array}
\right)
\end{equation}

\noindent
and coincides with the correspondent monodromy in the Seiberg-Witten model.
The further arguments completely repeat the derivation of Seiberg-Witten
exact result by the method, proposed in \cite{bilal}. Finally, we conclude,
that the functions $a(u)$ and $a_D(u)$ coincide with the Seiberg-Witten
solution (\ref{sw}).

Now it is quite clear, that the substitution $z=a^{-4}$ was made in order
to obtain the correct structure of the expansion (\ref{stru}). The
uniqueness of the solution (with the condition $\mbox{Re}\, f < 0$) can
be proven similar to \cite{unique}.


\end{document}